\documentclass[aps,english,prl,amsmath,amsfonts,amssymb,superscriptaddress,twocolumn,showpacs,citeautoscript]{revtex4-1}
\usepackage[T1]{fontenc} \usepackage[latin9]{inputenc}
\usepackage{amsmath} 
\usepackage{amssymb}
\usepackage{wasysym} 
\usepackage{esint} 
\usepackage{graphicx}
\usepackage{times} 
\usepackage{nicefrac} 
\usepackage{appendix}
\usepackage{textcase} 
\usepackage{subfigure} 
\usepackage{empheq}
\usepackage{color,soul} 
\usepackage{leftidx} 
\usepackage{makecell}
\usepackage{stackrel} 
\makeatletter

\newcommand{\Tr}{\text{Tr}}
\renewcommand{\Re}{\text{Re}}
\renewcommand{\Im}{\text{Im}}

\renewcommand{\eqref}[1]{(\ref{eq:#1})}
\newcommand{\Eqref}[1]{Equation~(\ref{eq:#1})}

\begin{document}
\title{Material Scaling and Frequency-Selective Enhancement of Near-Field Radiative Heat Transfer for Lossy Metals in Two Dimensions via Inverse Design}

\author{Weiliang Jin}
\affiliation{Department of Electrical Engineering, Princeton University, Princeton, NJ, 08544}

\author{Sean Molesky}
\affiliation{Department of Electrical Engineering, Princeton University, Princeton, NJ, 08544}

\author{Zin Lin}
\affiliation{John A. Paulson School of Engineering and Applied Sciences Harvard University, Cambridge, MA, 02138}

\author{Alejandro W. Rodriguez}
\affiliation{Department of Electrical Engineering, Princeton University, Princeton, NJ, 08544}
\email{arod@princeton.edu} 

\begin{abstract}
\noindent
The super-Planckian features of radiative heat transfer in the
near-field are known to depend strongly on both material and geometric
design properties.  However, the relative importance and interplay of
these two facets, and the degree to which they can be used to
ultimately control energy flow, remains an open question.  Recently
derived bounds suggest that enhancements as large as
$|\chi|^4 \lambda^{2} / \left(\left(4\pi\right)^{2}
  \Im\left[\chi\right]^{2} d^{2}\right)$
are possible between extended structures (compared to blackbody); but
neither geometries reaching this bound, nor designs revealing the
predicted material ($\chi$) scaling, have been previously
reported. Here, exploiting inverse techniques, in combination with
fast computational approaches enabled by the low-rank properties of
elliptic operators for disjoint bodies, we investigate this relation
between material and geometry on an enlarged space
structures. Crucially, we find that the material proportionality given
above does indeed emerge in realistic structures.  In reaching this
result, we also show that (in two dimensions) lossy metals such as
tungsten, typically considered to be poor candidate materials for
strongly enhancing heat transfer in the near infrared, can be
structured to selectively realize flux rates that come within $50\%$
of those exhibited by an ideal pair of resonant lossless metals for
separations as small as $2\%$ of a tunable design wavelength.
\end{abstract}
\maketitle

Radiative heat transfer (RHT) between objects separated by near-field
distances (on the order or shorter than the thermal wavelength)
exhibits a number of remarkable features. Primarily, evanescent
contributions, absent in the far-field, cause the rate of heat
exchange to scale inversely with separation down to nanometer scales,
leading to flux rates many orders of magnitude larger than those
predicted by the Stefan--Boltzmann (blackbody)
law~\cite{kittel2005near,kim2015radiative,song2015enhancement,st2016near}.
Further, this increased flux can be enhanced and controlled by
resonant electromagnetic surface
modes~\cite{pendry1999radiative,carminati1999near,volokitin2001radiative,babuty2013blackbody},
allowing heat to be concentrated into narrow and designable spectral
bandwidths. These properties, in principle, provide a means of
significantly improving the degree to which heat can be manipulated
compared to thermal conduction, leading to the consideration of
applications and devices based on near-field RHT in various contexts,
with proposals ranging from thermophotovoltaics energy
capture~\cite{narayanaswamy2003surface,laroche2006near,park2008performance,ilic2012overcoming,karalis2017transparent,st2017hot}
to high-density heat
management~\cite{guha2012near,yang2013radiation,khandekar2017near},
and heat assisted magnetic
recording~\cite{challener2009heat,Bhargava}.

Yet, a concrete understanding of what can be accomplished with
near-field RHT remains elusive. Simple high-symmetry structures where
analytic solutions are possible provide valuable insight, but appear
to be far from ideal. In particular, the most well-studied platform
for implementing selective RHT
enhancement~\cite{volokitin2007near,iizuka2015analytical}, involving
parallel metal
plates~\cite{domoto1970experimental,kralik2012strong,ijiro2015near,lang2017dynamic}
supporting surface resonances (plasmon or phonon
polaritons)~\cite{shen2009surface,luo2013nanoscale}, has critical
deficiencies.  First, as dictated by the Planck distribution, there is
a natural wavelength scale for observing significant thermal radiative
effects near ambient temperatures that spans the near to mid infrared
(1 to $10\;\mu$m)
spectrum~\cite{basu2007microscale,rephaeli2009absorber,molesky2013high}. Typical
(low-loss) optical materials do not support polariton resonances at
these wavelengths, and often lack sufficient thermal stability to
withstand longterm
operation~\cite{nagpal2011fabrication,rinnerbauer2013high,dyachenko2016controlling}.
Second, the tightest known limits of RHT between extended structures,
recently derived using energy conservation and reciprocity
arguments~\cite{miller2015shape}, reveal that both practical material
($|\chi|^2/\Im\,\chi$) and geometric ($\lambda/d$) factors seemingly
enable enhancements relative to blackbody emission as large as
\begin{equation}
  \mathcal{F}_\mathrm{limit} \leq \left(\frac{1}{4\pi}\frac{\lambda}{d}\frac{|\chi|^2}{\Im[\chi]}\right)^{2},
\label{eq:enerBounds} 
\end{equation}
orders of magnitude larger than what is achievable with ideal planar
media, suggesting that dramatic improvements are possible through the
use of nanostructured surfaces~\cite{miller2015shape}. (In this
expression, $d$ stands for separation, and $\chi$ material
susceptibility, assumed to be the same in both bodies.) Moreover, the
$|\chi|^4/\Im\left[\chi\right]^2$ scaling in \eqref{enerBounds}
indicates that materials exhibiting strong metallic response, far from
the typical planar surface--plasmon polariton (SPP) condition
$\Re\left[ \chi\right]=-2$, have much greater potential for
enhancement. To date, however, this behaviour has not been observed,
and tested geometries employing non-resonant
materials~\cite{fernandez2017enhancing,dai2017ultrabroadband,guerout2012enhanced,lussange2012radiative,liu2013broadband}
have yet to surpass the optimal rates associated with planar bodies.

In this article, we provide direct evidence of this
material scaling enhancement in periodic systems.  Building on our
earlier examination of RHT between multilayer
bodies~\cite{weiliang2017OE}, we now employ inverse
design~\cite{molesky2018outlook} to investigate RHT between generalized
two-dimensional gratings (restricting the analysis to realistic materials and fabricable structures). Focusing on lossy metals far from the SPP
condition at infrared wavelengths, we observe that while $\mathcal{F}$
does not approach \eqref{enerBounds}, the predicted material scaling
is nevertheless present.  For the specific example of tungsten (W), we
also find a grating geometry that selectively achieves $50\%$ of the
RHT of a pair of ideal, lossless metal plates satisfying the SPP
condition, for separations as small as $2\%$ of a tunable design
wavelength.  These predictions represent RHT enhancements of nearly
two orders of magnitude compared to corresponding planar objects,
confirming the potential of even relatively simple structuring for
selectively enhancing RHT.

The application of inverse design to selective RHT enhancement between
extended structures is complicated in several ways. First, near-field
RHT is controlled by evanescent electromagnetic fluctuations. The
large density of these states makes it challenging to apply
traditional resonant nanophotonic strategies for enhancing far-field
emission over narrow spectral
windows~\cite{pralle2002photonic,de2012conversion}. Moreover, the
characteristically large field amplitudes and sub-wavelength features
of evanescent states make them sensitive to small variations in
structural and material properties~\cite{liu2010infrared}, and
correspondingly accurate modeling of RHT requires fine numerical
resolution~\cite{rodriguez2011frequency,rodriguez2012fluctuating,otey2014fluctuational,messina2017radiative}.)
Second, unlike the far field, RHT can not be decomposed into
approximately equivalent independent subproblems. Alterations in the
structure of any one object affects the response of the entire system,
meaning that the scattering properties of all bodies must be
controlled simultaneously. Finally, Maxwell's equations depend
nonlinearly on the dielectric properties and shapes of all bodies,
making the optimization non-convex~\cite{boyd2004convex} and any a
priori guarantee of globally optimal solutions impossible.

Consequently, tractable general approaches for simply calculating
near-field RHT have only recently been
realized~\cite{rodriguez2011frequency,rodriguez2012fluctuating,otey2014fluctuational,sheila2016surface},
and nearly all previously studied geometries have been designed via
trial-and-error approaches exploiting brute-force search over a
handful of high-symmetry design parameters~\cite{liu2015outlook}.
Beginning with bulk
metamaterials~\cite{biehs2011nanoscale,guo2012broadband,liu2013broadband,liu2013near,liu2014nearmeta,shi2015near,didari2015near},
thin
films~\cite{ben2009near,francoeur2010spectral,basu2011maximum,van2011phonon,pendharker2017thermal},
plasmonic
materials~\cite{naik2013alternative,pendharker2017thermal,liu2014nearmeta,sheila2016surface,chalabi2016focused},
and more recently, metallic
metasurfaces~\cite{liu2015near,fernandez2017enhancing,zheng2018spectral}
and
gratings~\cite{lussange2012radiative,LiuAPL14,Dai1,Liu2015enhanced,chalabi2016focused},
selective RHT enhancement has primarily been achieved by tuning the
permittivity response, either real or effective, to create or mimic
surface resonances~\footnote{There is a considerable body of work in
  this area. The accuracy of effective medium approximations has been
  explored for both one- and two-dimensional grating structures, with
  and without dielectric filling~\cite{Dai2,Dai3}.}.
(Notably, a recent silicon metasurface
design~\cite{fernandez2017enhancing} was predicted to have a larger
integrated RHT than planar SiO$_2$, which exhibits low-loss surface
phonon polaritons, down to gap distances of $10\;\text{nm}$.)  Other
similarly high symmetry approaches have sought to increase the
photonic density of states by exploiting interference
(hybridization~\cite{prodan2003hybridization}) among the localized
plasmons of individual
nanostrutures~\cite{biehs2010mesoscopic}. Building from simple shapes,
tunable RHT rates have been demonstrated in nanobeam (triangluar,
ellipsoidal, and rectangular unit cells) and nanoantenna arrays
exploiting both Mie~\cite{chalabi2015effect} and
Fano~\cite{baldwin2014thermal,perez2017fano} resonances. Although
conceptually promising, these approaches have been found to have
diminishing returns at small separations (relative to the thermal
wavelength). 


In the case of periodic gratings, RHT is 
\begin{align}
  \Phi(\omega)=\frac{1}{2\pi}\int \frac{\mathrm{d}^n\mathbf{k}}{(2\pi)^n}\;\mathcal{T}\left(\omega ,\textbf{k}\right),
\label{eq:T}
\end{align}
where the integration is carried out over Bloch-vectors $\mathbf{k}$
in the first Brillouin zone (BZ), and the scattering properties of the
structures are captured by the transfer function
$\mathcal{T}\left(\omega ,\textbf{k}\right)$ described by
\eqref{GreenHeat}.  To maximize the transfer function, both the
density and coupling efficiency of the participating
states~\cite{molesky2015ideal,liu2017resonant} must be made as large
as possible at all $\textbf{k}$.  A similar problem arises in the
design of far-field
emitters~\cite{wangLDOS,Ganapati,diem2009wide,liu2010infrared,hedayati2011design,nguyen2017control},
where resonant structures are often intuitively designed to match the
absorption and coupling rates of a wide range of externally excited
states, leading to a complete suppression of scattered fields.  But
the task here is more complicated, as the electromagnetic field must
be regulated over a much larger (evanescently coupled) range. For
intuition based structures with a few tunable parameters, there does
not seem to be enough design freedom to reach this level of control,
with the range of rate-matched states occurring in current
designs~\cite{dai2017ultrabroadband} falling short of those achievable
in planar high index dielectrics ($n>3$). Without a viable means of
addressing these questions, we turn to inverse
techniques~\cite{molesky2018outlook}.



As an initial step towards the broader development of this area, and for computational convenience, we focus
exclusively on two-dimensional gratings.  This choice has major
consequences for the underlying physical processes.  Particularly, in
moving from three to two dimensions, the geometric and material
scaling of the density of states decreases and as a result, the
maximum RHT rate between two ideal planar metals,
$\mathrm{Re}\left[\chi\left(\omega\right)\right]=-2$, becomes finite
in the limit of vanishing loss~\cite{miller2015shape},
\begin{equation}
  \mathcal{F}^{2D}_\text{pl}\left(\omega\right) = \frac{1}{2\pi}\frac{\lambda}{d},
  \label{eq:uniLimit}
\end{equation}
$\mathcal{F}(\omega)=\Phi(\omega)/\Phi_0(\omega)$, and
$\Phi_0(\omega) = \omega/\pi^2 c \;\;(\omega^2/4\pi^{2} c^2)$ the
spectral emission rate per unit area of a two (three) dimensional
planar blackbody. In contrast, $\mathcal{F}^{3D}_\text{pl}\left(\omega\right) =
\lambda^2\ln\left(2/\Im[\chi(\omega)]\right)/\left(2\pi^2d^2\right)$
exhibits stronger geometric and material enhancement
factors. Consequently, achieving a strong material response
$\chi\left(\omega\right)$ at the desired frequency window along with
broadband rate-matching through nanostructuring is expected to have
more significant impact in three dimensions. As  
\eqref{uniLimit} is useful standard for comparing the efficacy of any given
design it will be used as a normalization throughout the
remainder of the manuscript.

Keeping these considerations in mind, we
now describe a computational method that allows fast computations of
RHT between arbitrarily shaped gratings of period $\Lambda$, separated
by a vacuum gap $d$.
\begin{figure}[t!]
  \centering
  \includegraphics{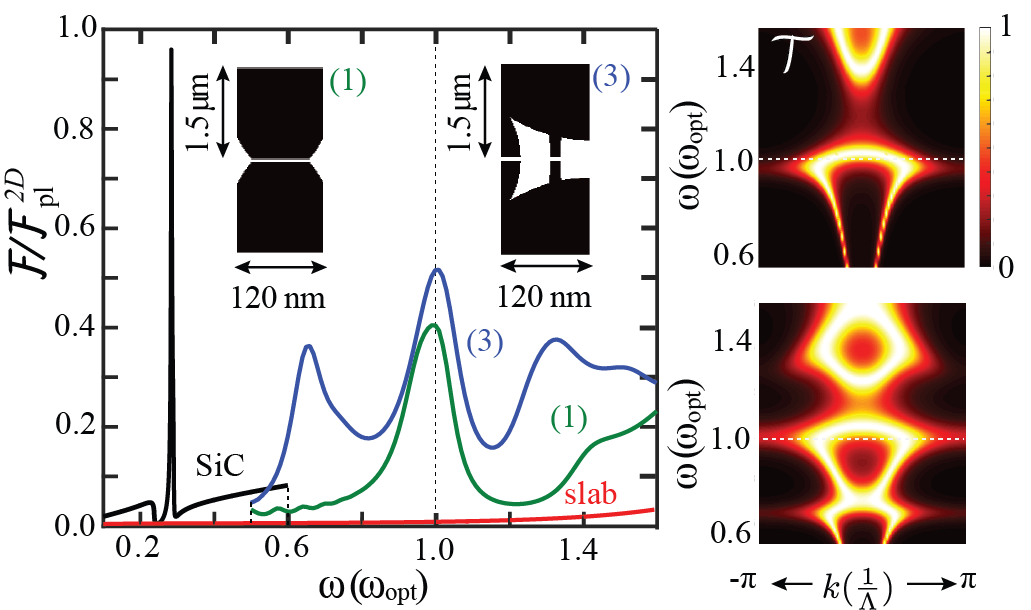}
  \vspace{-10 pt}
  \caption{\textbf{Inverse design of selective near-field heat
      transfer between periodic tungsten gratings}: Near-field RHT
    enhancement for inverse-designed tungsten gratings $\mathcal{F}$,
    along with that of planar silicon carbide for comparison, relative
    to the enhancement achieved with ideal (lossless) planar metals,
    $\mathcal{F}^{2D}_\text{pl}$ given in \eqref{uniLimit}, with
    respect to frequency $\omega$ (left).  Profiles of the structures
    are displayed as insets. Color plots depicting the $k$-dependent
    transfer functions $\mathcal{T}(\omega,\mathbf{k})$ over the
    chosen frequency range are shown on the right. The depicted
    optimization proceeds from an unstructured planar system (0) to
    structured gratings (N) by successively introducing additional
    ellipsoidal degrees of freedom ``N'' to the design
    space. Performance is qualified by the magnitude of RHT at a
    single design frequency $\omega_\text{opt} = 2\pi c / 3\;\mu$m
    ($\lambda_\text{opt}=3\;\mu$m), where tungsten behaves as a highly
    lossy metal far from the planar surface plasmon resonance. The gap
    separation and the period of gratings are
    $d(\Lambda)=0.02\;(0.04)\;\lambda_\text{opt}$, respectively. Even
    in the case of a single ellipsoid, producing a relative simple
    grating, heat transfer is enhanced by a factor of 40, while the
    more complicated design (3) is observed to come within $50\%$ of
    the ideal planar limit.}
  \vspace{0 pt}
\end{figure}
Within fluctuational electrodynamics, the calculation of RHT consist
of determining the absorbed power within a body $B$,
$\Phi\left(\omega\right)=\frac{1}{2}\omega\varepsilon_o\int_{V_{B}}d\textbf{r}'\;\Im\left[\chi\left(\textbf{r}',\omega\right)\right]\;\left<|\mathbf{E}\left(\textbf{r}',\omega\right)|^2
\right>$, resulting from thermally excited current sources originating
within a different body $A$. Given a discretized computational grid
and assuming local media, these sources obey the
fluctuation--dissipation relation~\cite{rytov1988principles},
\begin{equation}
\langle j_{\gamma,i}^{*}\;j_{\beta,j}\rangle=\frac{4\omega\epsilon_{o}}{\pi}\;\delta_{ij}\;\delta_{\gamma\beta}\;\Im[\chi_{i}\left(\omega\right)]\Theta\left(\omega,T_{i}\right).
\label{eq:corrRel}
\end{equation}
Here, $\Theta(\omega,T)=\hbar\omega/(\mathrm{e}^{\hbar\omega/k_BT}-1)$
is the Planck function, $\left<\ldots\right>$ a thermal ensemble
average, $\left\{i,j\right\}$ the index of a given location or pixel
within the computational grid, and $\{\gamma,\beta\}=\{x,y,z\}$ the
vector polarizations. \Eqref{corrRel} can be used in conjunction with
knowledge of the electric Green's function $\mathcal{G}$ of the
system~\cite{kong1975theory}, to obtain,
\begin{align}
\Phi\left(\omega\right)=&\frac{\omega^4\Theta\left(\omega,T\right)}{2\pi c^4}\sum_{\beta,\gamma}\; \int_{V_{A}}d\textbf{r}\int_{V_{B}}d\textbf{r}'\;\nonumber \\
&\Im\left[\chi\left(\textbf{r},\omega\right)\right] \Im\left[\chi\left(\textbf{r}',\omega\right)\right] \left| \mathcal{G}_{\beta\gamma}\left(\textbf{r}',\textbf{r},\omega\right) \right|^2.
\end{align}
Writing this in matrix form, with superscripts denoting projections
onto the respective body, and $\mathcal{G}$ denoting the matrix form of
the electric Green's function, it follows that RHT can be written as
a Frobenius norm,
\begin{align}
  & \Phi\left(\omega\right)= \frac{\omega^4\;\Theta\left(\omega,T\right)}{2\pi c^4}\; ||\sqrt{\Im\left[\chi^{_{A}}\right]}\mathcal{G}^{_{AB}} \sqrt{\Im\left[\chi^{_{B}}\right]}||_{F}^{2}. 
  \label{eq:GreenHeat}
\end{align}
The main challenge in evaluating \eqref{GreenHeat} lies in the need to
repeatedly evaluate and multiply $\mathcal{G}^{_{AB}}$, the inverse of
a sparse matrix. Direct application of either
sparse-direct~\cite{li2003superlu_dist} or iterative
solvers~\cite{plessix2007helmholtz} would demand extraordinary
computational resources, especially in three dimensions.  In
particular, without additional simplifications, within a particular
numerical discretization, the number of computations at each iteration
of an optimization required is at least the rank of the matrix
$\sqrt{\Im\left[\chi^{_{B}}\right]}$, or three times the number pixels
in $B$ (polarizations).  However, because $\mathcal{G}^{_{AB}}$ does
not describe fields created by current sources within the same body
(but only disjoint bodies), it admits a low-rank
approximation~\cite{gatto2015preconditioner}. Hence, \eqref{GreenHeat}
can be well approximated by a singular value decomposition of the
matrix
$\mathcal{Z}^{_{AB}} =
\sqrt{\Im\left[\chi^{_{A}}\right]}\mathcal{G}^{_{AB}}
\sqrt{\Im\left[\chi^{_{B}}\right]}$,
\begin{align}
  & \Phi\left(\omega\right)= \frac{\omega^{4}
    \;\Theta\left(\omega,T\right)}{2\pi c^4}\;
  \sum_{i}|\sigma_{i}|^{2}, \label{SVDHeat}
\end{align}
requiring only a small set of singular values
$\left\{\sigma_{i}\right\}$.  Applying the fast randomized SVD
algorithm~\cite{halko2011finding}, detailed in Appendix A, we find
that typically no more than $8$ singular values are needed to reach an
error estimate better than $1\cdot10^{-3}$, reducing the number of
required matrix solves to $\lesssim 16$.  (As derived in Appendix B,
this trace formulation also enables fast gradient computations via the
adjoint method~\cite{molesky2018outlook}.) 

The inverse problem is then to maximize $\sum_{i}|\sigma_{i}|^{2}$
with respect to variations in $\chi$.  Such an optimization can be
carried out in the framework of topology optimization using the
adjoint method~\cite{molesky2018outlook}, allowing a huge range of
design parameters (each pixel within the optimization domain).  We
find, however, that local, gradient-based optimization leads to slow
convergence to fabricable structures and comparatively suboptimal
designs.  To avoid these difficulties, we instead considered a range
of shape optimizations~\cite{molesky2018outlook}.  While limiting the
space of discoverable structures, this choice allows for application
of statistical Bayesian
algorithms~\cite{caers2006probability,martinez2014bayesopt} in
combination with fast, gradient-based optimization.  Specifically, the
susceptibility profile over the periodic computational domain is
described by the product $\chi_i=\bar{\chi}\prod_\alpha
f_\alpha(\mathbf{x}_i;\{p_\alpha\},\zeta_i)$, where $\bar{\chi}$
denotes the susceptibility of the metal, and each $f_{\alpha}$ is a
shape function characterized by geometric parameters $\{p_\alpha\}$.
For improved convergence, $f_{\alpha}$ also contains a smoothing
kernel, allowing gradual variations between metal and non-metal
regions.  To obtain a binary structure, the $\left\{\zeta_i\right\}$
smoothing parameters are reduced at each successive iteration of the
optimization until the shape functions are piece-wise
constant~\cite{tsuji2008design}. In what follows, this procedure,
implemented with a simple 2d FDFD Maxwell solver~\cite{xu2003finite},
is applied to selectively enhance RHT at the thermal wavelength
$\lambda_\text{opt} = 3\;\mu$m corresponding to peak emission at
$T=1000$~K. The surface--surface separation between the two bodies is
fixed to be $d=0.02\;\lambda_\text{opt}$ (deeply in the sub wavelength
regime).
\begin{figure}[t!]
  \centering \includegraphics{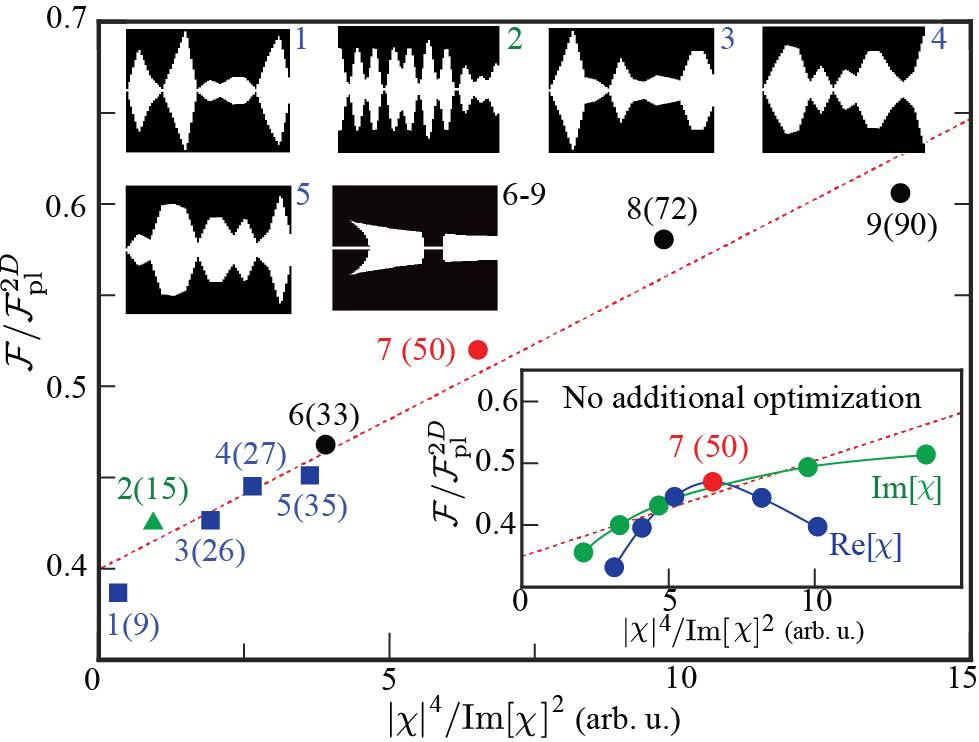}
  \vspace{0 pt}
  \caption{\textbf{Observation of material scaling in shape
      optimization}: The figure highlights the material trend in
    enhancement $\mathcal{F}/\mathcal{F}^{2D}_\text{pl}$ found by
    optimizing grating geometries over a wide range of material
    susceptibilities $\chi$.  Quantities displayed in parenthesis
    correspond to the relative enhancement of a given grating design
    compared to a planar geometry of the same material.  Various shape
    parameterizations such as polylines, Fourier curves, and
    ellipsoids marked as squares, triangles, and circles are
    considered. The susceptibility values are chosen to vary along
    either the real or imaginary axis: gratings $\{1,2,4\}$ correspond
    to $\chi=\{-76,-101,-151\}+50i$, $5$ to $\chi=-151+40i$, $\{3,7\}$
    to $\chi=\{-121,169.5\}+37.3i$, and $\{6\rightarrow9\}$ to
    $\chi=-169.5+\{50,37.3, 30,25\}i$.  Results for tungsten are
    colored red. The upper inset illustrates the unit cell of
    correspondingly numbered grating. (There is no visible difference
    for gratings 6-9 and so only one of these is shown.) The lower
    inset depicts the RHT of tungsten grating 7 when the
    susceptibility is varied without altering the structure. The
    dashed line represents the susceptibility scaling predicted by
    recently derived limits on
    RHT~\cite{miller2015shape}.}  \vspace{-10 pt}
\end{figure}

We begin by considering tungsten gratings, which at
$\lambda_{\text{opt}}$ exhibits a highly metallic response $\chi \approx
-170 + 37i$ far from the planar SPP condition. Figure~1 depicts the
spectral enhancement factor $\mathcal{F}/\mathcal{F}^{2D}_\text{pl}$
for both unstructured plates ``$(0)$'' and optimized gratings
``$(N)$'' obtained by successively increasing the number of
(ellipsoidal) shapes allowed in each unit cell, $N$. The spectra of
the two optimized gratings, illustrated as insets~\footnote{Full
  geometric characterizations of these structures and those presented
  later in the text are available upon request.}, both peak at
$\lambda_{\text{opt}}$ (black dashed line), with magnitudes
$\mathcal{F}/\mathcal{F}^{2D}_\text{pl}=\{0.40,0.53\}$ increasing
with the number of ellipsoids, $\{1,3\}$\footnote{The optimizations
  involving four elliptical bodies usually require $\approx 10^{2}$
  iterations to reach convergence; roughly $200$ hours of total
  computation time for each structure.}.  On the one hand,
enhancements of this magnitude for lossy metals,
$|\chi|/\mathrm{Im}\left(\chi\right) \gtrsim 1$, are considered
challenging~\cite{raether1988surface} at small separations, $d\ll
\lambda_{\text{opt}}$.  On the other hand, large RHT is known to be
possible in ultra-thin films through the interference of coupled
SPPs~\cite{guo2014thermal}. However, to reach the magnitudes obtained
here, unrealistically small thicknesses $\lesssim 1$~nm are needed.
In contrast, no feature in the gratings of Fig.~1 is smaller than
$10$~nm.  Notably, although the optimization is carried out at a
single frequency, the discovered enhancement peaks are always
broadband due to the high level of material absorption.
Consequently, the frequency-integrated RHT at $T=1000$~K exhibited by
grating $(3)$ is found to be roughly $10\%$ larger than that of two
planar silicon carbide (black line), a low-loss polaritonic material.
To explain this enhancement, Fig.~1 (right) examines the transfer
function $\mathcal{T}(\omega,k)$ versus frequency and wavenumber $k$.
In moving from (1) to (3), the color plot demonstrates (frequency
axis) that additional modes are successively created and pushed
towards $\omega_{\mathrm{opt}}$, enhancing the density of
states. Owing to the large size of the BZ (small periodicity
$\Lambda=0.04\;\lambda_{\text{opt}}$), we find that the range of rate
matching achieved here is considerably larger than that observed in
previously examined grating structures~\cite{Dai1}.

Another key finding is depicted in Fig~2, which plots RHT enhancement
for representative optimizations~\footnote{The structures with the
  best performance features over all trials with the same
  susceptibility.}  across an array of material and geometry
combinations, as a function of the material scaling factor
$|\chi|^{4}/\Im\left[\chi\right]^2$ of \eqref{enerBounds}.  Three
different classes of design are explored: collections of ellipsoids
(circles), single polyline interfaces (squares), and Fourier curves
(triangles).  Uniformly, every one of these structures is observed to
enhance RHT by at least an order of magnitude compared to the
corresponding planar systems and within factors of unity of the ideal
planar bound, \eqref{uniLimit}. Regardless of the particular
parameterization, over the range of examined $\Re[\chi]$ and
$\Im[\chi]$, a clear linear trend in $\mathcal{F}$ (red dotted line)
as a function of $|\chi|^{4}/\Im\left[\chi\right]^2$ is observed.
This linear scaling becomes increasingly difficult to observe at
larger values of $\chi$, where larger resolutions are needed to
accurately capture resonant behavior and the optimization requires
increasingly larger number of iterations to find structures along the
fit line. As should be expected, based on the fact that this behavior
has not been previously reported, material scaling consistent with
\eqref{enerBounds} is seen only for gratings optimized specifically
for each particular value of $\chi(\omega_\mathrm{opt})$.  For a fixed
geometry, Fig.~2 (insets), varying either $\Re[\chi]$ or $\Im[\chi]$
shifts the resonance away from $\omega_\mathrm{opt}$ and diminishes
RHT.  The appearance of this linear trend indicates that aspects of
the arguments leading to the energy bounds in\cite{miller2015shape}
are coming into play. However, the fact that a substantial increase in
available design space has failed to significantly bridge the
magnitude gap suggests that approaching these bounds (at least in 2d)
may prove challenging with practical structures. Conversely, it should
be emphasized that our results do not preclude the existence of
structures with much larger enhancements. First, although the design
space we have investigated is substantially larger than previous work,
it is still relatively limited.  Second, the complexity of the
strutures and degree of enhancement are limited by the spatial
resolution of the chosen discretization, $0.0005\;\lambda_\text{opt}$
(1/40th of the gap size).


To summarize, in investigating potential radiative heat transfer
enhancements through inverse design, we have found evidence supporting
the material scaling recently predicted by shape-indepedent
bounds~\cite{miller2015shape}, a feature that to our knowledge had yet
to be confirmed.  While the observed heat transfer rates are still far
from matching the magnitudes predicted by general bounds, we have
found that RHT rates between fabricable tungsten gratings (a highly
lossy metal), for subwavelength gap separations as small as $2\%$ of
the design wavelength, can selectively approach $50\%$ of the rate
achieved by ideal planar materials (lossless metals satisfying the SPP
condition) in the infrared. The results represent nearly two orders of
magnitude greater RHT rates in structured compared to planar
materials. It remains to be seen to what degree similar strategies
might enhance RHT in three dimensions, where the photonic density of
states is larger and its associated dependence on material losses
significantly stronger.

This work was supported by the National Science Foundation under Grant
No. DMR-1454836, Grant No. DMR 1420541, and Award EFMA-1640986; and
the National Science and Research Council of Canada under
PDF-502958-2017.

\appendix{\noindent \textbf{Appendix A: Approximate
    Singular Value Decomposition}}

In this appendix, we sketch how the low-rank nature of the RHT matrix
$\mathcal{Z}^{_{AB}} =
\sqrt{\Im\left[\chi^{_{A}}\right]}\mathcal{G}^{_{AB}}
\sqrt{\Im\left[\chi^{_{B}}\right]}$
entering (6) in the main text allows application of fast randomized
singular value decompositions (SVD), greatly speeding up calculations
of RHT. We begin by splitting the problem into two domains, referred
to as bodies $A$ and $B$, with associated superscripts denoting
projection. It will be assumed that sources occur in body $A$ and that
the fields of interest lie only in body $B$. (From Lorentz
reciprocity, the Green function must be symmetric under the exchange
of observation and source positions, and so there is no loss of
generality in either choice.)  The algorithm, described in detail in
\cite{halko2011finding}, proceeds iteratively as follows.

Draw a random Gaussian distributed current vector
$\tilde{\textbf{j}}_{i+1}$ with values in body $A$. Let
$\tilde{\Omega}_{i}^{_A}$ denote the set of all previously drawn
randomly Gaussian distributed current vectors (vertically concatenated
to form a matrix). Solve Maxwell equation's to obtain the associated
field,
$\tilde{\mathbf{E}}_{i+1}=\mathcal{G}\tilde{\textbf{j}}_{i+1}$. Let
$\tilde{O}^{_B}_{i}$ denote the set of all previously computed,
orthonormalized, electromagnetic field vectors, and
$\tilde{Y}^{_B}_{i}$ the set of all previous electromagnetic fields as
originally calculated. Compute
$\tilde{\textbf{o}}^{u\;_{B}}_{i+1}=\left(I-\tilde{O}^{_B}_{i}\tilde{O}^{\dagger_{B}}_{i}\right)\tilde{\mathbf{E}}_{i+1}$,
and normalize the result
$\tilde{\textbf{o}}^{_B}_{i+1}=\tilde{\textbf{o}}^{u_B}_{i+1}/\langle\tilde{\textbf{o}}^{u_B}_{i+1}|\tilde{\textbf{o}}^{u_B}_{i+1}\rangle$,
recording the value of $\epsilon_{i+1} =
\langle\tilde{\textbf{o}}^{u_B}_{i+1}|\tilde{\textbf{o}}^{u_B}_{i+1}\rangle$.
Concatenate $\tilde{\textbf{o}}^{_B}_{i+1}$ onto $\tilde{O}^{_B}_{i}$
and $\tilde{\mathbf{E}}_{i+1}$ onto $\tilde{Y}^{_B}_{i}$, producing
$\tilde{O}^{_B}_{i+1}$ and $\tilde{Y}^{_B}_{i+1}$. Similarly, draw
random currents $\tilde{\textbf{j}}_{i+1}$ in body $B$ and repeat the
previous procedure, producing $\tilde{O}^{_A}_{i+1}$ and
$\tilde{Y}^{_A}_{i+1}$.  The iterations are stopped when both
$\epsilon_{i+1}$ and $\epsilon^{\dagger}_{i+1}$ (for both $A$ and $B$
calculations) are smaller than a prescribed singular value tolerance.

Given the above matrices, a low-rank approximation of the SVD of
$\mathcal{Z}^{_{AB}}$ is obtained by expanding it onto the basis
functions $\tilde{O}^{_A}$ and $\tilde{O}^{_B}$. From random matrix
theory~\cite{halko2011finding}, given that $\epsilon_{i+1}$ is small
for a given number of successive iterations, this basis approximately
spans the domain and range of the matrix. It follows from Maxwell's
equations that $\tilde{Y}^{_A}=\mathcal{G}^{_{AB}}\tilde{\Omega}^{_B}$
and hence,
\begin{equation}
\tilde{O}^{_A\dagger}\mathcal{Z}^{_{AB}}\tilde{O}^{_B}\tilde{O}^{_B\dagger}\sqrt{\Im[\chi^{_{B}}]}^{-1}\tilde{\Omega}^{_B}\approx\tilde{O}^{_A\dagger}\sqrt{\Im[\chi^{_{A}}]}\tilde{Y}^{_A}.\nonumber
\end{equation}
Multiplication by the inverse of
$\tilde{O}^{_B\dagger}\sqrt{\Im[\chi^{_{B}}]}^{-1}\tilde{\Omega}^{_B}$ then produces a low-dimension $k\times k$ matrix on the right hand side, amenable to standard SVD at minimal cost,
\begin{align}
  &\tilde{O}^{_A\dagger}\mathcal{Z}^{_{AB}}\tilde{O}^{_B} \approx \underbrace{ \tilde{O}^{_A\dagger}\sqrt{\Im[\chi^{_{A}}]}\tilde{Y}^{_A}\left(\tilde{O}^{_B\dagger}\sqrt{\Im[\chi^{_{B}}]}^{-1}\tilde{\Omega}^{_B}\right)^{-1}}_{U\Sigma V^{\dagger}}.
\end{align}
The singular value approximation of $\mathcal{Z}^{_{AB}}$ is then derived from the small $k\times k$ matrices $U$,
$\Sigma$, and $V$,
\begin{align}
    \mathcal{Z}^{_{AB}}&\approx\tilde{O}^{_A}U\Sigma V^{\dagger}\tilde{O}^{_B\dagger}\nonumber\\
    &=\left(\tilde{O}^{_A}U  \right)\Sigma \left( \tilde{O}^{_B}V \right)^{\dagger}
\end{align}
Analysis of the convergence and performance properties of similar
algorithms have been previously produced in~\cite{halko2011finding}.

\appendix{\noindent\textbf{Appendix B: Fast Gradient Adjoints}}

To exploit gradient-based optimization, knowledge of $\partial \Phi
/\left( \partial p_{\alpha}\right)$ is required for each optimization
parameter $p_{\alpha}$. Letting $\partial_{\alpha}$ denote a partial
derivative with respect to $p_\alpha$, and retaining the notation of
the main text and Appendix A, 
\begin{align}
\partial_{\alpha} \Phi 
&=\frac{\omega^{4}\Theta}{2\pi c^{4}}\partial_{\alpha}\Tr\left[\Im\left[\chi^{_{B}}\right]\mathcal{G}^{_{AB}\dagger}\Im\left[\chi^{_{A}}\right]\mathcal{G}^{_{AB}}\right].
\end{align}
Using the symmetry of $\chi^{_{A}}$ and $\chi^{_{B}}$, along with the
usual cyclic properties of the trace,
\begin{align}
  &\Tr\left[\Im\left[\chi^{_{B}}\right]\partial_{\alpha}\mathcal{G}^{_{AB}\dagger}\Im\left[\chi^{_{A}}\right]\mathcal{G}^{_{AB}}\right]= \nonumber \\
  &\Tr\left[\mathcal{G}^{_{AB}\intercal}\Im\left[\chi^{_{A}}\right]\partial_{\alpha}\mathcal{G}^{_{AB*}}\Im\left[\chi^{_{B}}\right] \right] = \nonumber \\
  &\Tr\left[\Im\left[\chi^{_{B}}\right]\mathcal{G}^{_{AB}\intercal}\Im\left[\chi^{_{A}}\right]\partial_{\alpha}\mathcal{G}^{_{AB*}} \right],\nonumber
\end{align} 
one finds that the second and fourth terms in the partial derivative
expansion are complex conjugates. (Here, $\intercal$ denotes
transposition without complex conjugation.)  Given that $\mathcal{G} =
\left[\left(\nabla\times\nabla\times\right)-\omega^{2}\epsilon\right]^{-1}$
is the inverse Maxwell operator,
$$ \partial_{\alpha} \mathcal{G}^{_{AB}} =
\omega^{2}\;\mathcal{G}^{_A\cdot}\partial_{\alpha}\chi\;\mathcal{G}^{\cdot_{B}},$$
and so one finds:
\begin{align}
&\partial_{\alpha} \Phi =\frac{\omega^{4}\Theta}{2\pi c^{4}}\Tr\Bigg[\mathcal{G}^{_{AB}\dagger}\Im\left[\chi^{_{A}}\right]\mathcal{G}^{_{AB}}\Im\left[\partial_{\alpha}\chi^{_{B}}\right] +\nonumber \\
&\mathcal{G}^{_{AB}}\Im\left[\chi^{_{B}}\right]\mathcal{G}^{_{AB}\dagger}\Im\left[\partial_{\alpha}\chi^{_{A}}\right] + \nonumber \\
&2\omega^{2}\Re\left[\mathcal{G}^{\cdot_{B}}\Im\left[\chi^{_{B}}\right]\mathcal{G}^{_{AB}\dagger}\Im\left[\chi^{_{A}}\right]\mathcal{G}^{_{A}\cdot}\partial_{\alpha}\chi\right]\Bigg].
\label{eq:derExact}
\end{align}
The $\mathcal{G}^{_{AB}}$ matrices involved in \eqref{derExact} are
low rank, and can computed alongside $\mathcal{Z}^{_{AB}}$ at almost
no extra cost. Specifically, if $\sqrt{\Im\left[\chi^{_{A}}\right]}$
and $\sqrt{\Im\left[\chi^{_{B}}\right]}$ are excluded from the error
estimates $\epsilon_{i+1}$ and $\epsilon^{\dagger}_{i+1}$, one obtains
an approximation for $\mathcal{G}^{_{BA}}$ rather than
$\mathcal{Z}^{_{AB}}$. Both computations can be carried out at the
same time, using the same current sources and field solutions. If
these matrices have rank $k$, the parameter independent matrices of
\eqref{derExact} are solved in order $k$ computations. Hence,
determination of the gradient is essentially no more costly than the
determination of $\Phi$.

\end{document}